\documentstyle[12pt,aaspp4]{article}

\received{}
\revised{}
\accepted{}
\cpright{}{1996}

\journalid{}{}
\articleid{}{}
\paperid{}

\cpright{}{}
\ccc{}

\begin{document}

\title{The Interstellar Environment of Filled-Center Supernova Remnants 
III: The Crab Nebula}
\author{B.J. Wallace}
\author{T.L. Landecker}
\affil{Dominion Radio Astrophysical Observatory, Herzberg Institute of 
Astrophysics, National Research Council of Canada}
\authoraddr{P.~O. Box 248, Penticton, BC, Canada, V2A 6K3}
\author{P.M.W. Kalberla}
\affil{Radioastronomisches Institut der Universit\"at Bonn}
\authoraddr{Auf dem H\"{u}gel 71, Bonn D-53121, Germany}
\and
\author{A. R. Taylor}
\affil{Dept. Physics \& Astronomy, University of Calgary}
\authoraddr{Calgary, AB, Canada, T2N 1N4}

\begin{abstract}

The \ion{H}{1} environment of the Crab Nebula is 
investigated using $2.75'$ and $9'$ resolution data from (respectively) the DRAO
Synthesis, and 
Effelsberg 100~m, radio telescopes. No clear evidence for an 
interaction between the Crab and the surrounding \ion{H}{1} is found;
the Crab probably 
lies within the boundaries of a large scale, low-density void in the \ion{H}{1}
distribution. The presence of a wind-blown \ion{H}{1} bubble near the Crab is confirmed,
and it is suggested that the unidentified star which powers this bubble is
responsible for the stellar wind material detected along the line of sight 
towards the Crab.

\end{abstract}

\keywords{supernova remnants --- ISM: individual (Crab Nebula) --- ISM: bubbles --- ISM: general --- radio lines: ISM}

\section{Introduction}

The Crab Nebula is the prototype of a small subclass of supernova remnants
(SNRs) known as either ``Crab-like'', ``Filled-Center'' or ``plerionic''
SNRs. SNRs which fall into this class have centrally brightened radio 
morphologies, flat radio-spectral indices, high levels of linear polarization,
and lack an associated limb-brightened shell. 

It is not clear why Filled-Center (FC) SNRs do not have 
associated limb-brightened
shells. The cause could be intrinsic, i.e. there is a fundamental physical
difference between FC and other SNRs, or extrinsic, i.e. the lack of a shell 
is due to an environmental effect. The
extrinsic explanation was put forward by Chevalier~(1977), who
hypothesized that the Crab Nebula (and, by extension, all other FC SNRs) lie
in low-density regions of the ISM. In Chevalier's scenario the supernovae 
which produce FC SNRs are no different than other explosions which produce 
neutron stars. Such explosions cast off the outer envelope of the 
progenitor star at $\sim 10^4$~km/s, and the interaction between the resulting
blast wave and the ISM results in the typical SNR shell.
Assuming that the strength of the emission from the shell
is related to the density of the
ISM, the interaction between the fast moving ejecta and a sufficiently
low-density ISM might result in emission well below present detection
limits. FC SNRs would then be the result of SN explosions
which occur in low-density environments.

This paper is the third in a series in which the ISM around FC SNRs is probed 
to test Chevalier's hypothesis.
The first two papers (Wallace et al, 1997a, 1997b) have suggested that, 
contrary to expectations, at least two FC SNRs (G74.9+1.2, G63.7+1.1) do 
{\em not} lie in low-density regions of the ISM. Instead, clear indication
is seen for interaction between the SNRs and their surrounding ISM.
These results stand in contrast to previous
investigations of the ISM around the Crab Nebula itself (Romani et al, 1990;
Wallace et al, 1994) which suggest that the Crab {\it does}
lie in a low-density region 
of the ISM. However, interactions such as those
between G74.9+1.2 and G63.7+1.1 and
their surroundings would have been missed at the low resolution 
used for the existing studies of the Crab ($36'$); 
higher resolution observations
are thus required for definitive statements regarding the environment of 
the Crab Nebula.

In this paper we map the \ion{H}{1} distribution around the Crab Nebula at the 
highest resolution yet attempted. In our earlier paper (Wallace et al. 
1994; henceforth WLT94) 
we suggest that the Crab Nebula lies within a low-density void
which we referred to as ``bubble~2;'' in this paper we
use the Effelsberg 100-m radio telescope to image this feature at $9'$
resolution. 
In addition, the Dominion Radio Astrophysical Observatory (DRAO) Synthesis
Telescope is used to image the \ion{H}{1} environment immediately
around the Crab at $1' \times 3'$ 
resolution. The primary purpose of these data is to determine whether 
the Crab Nebula lies in a low-density void in the ISM. 

We also use these data to elaborate on the nature of the optical emission
detected by Murdin~(1994) and Fesen et al.~(1997). Murdin~(1994) reports
detecting faint H$\beta$ emission which he ascribes to the remnant stellar wind
from the Crab's progenitor. Fesen et al.~(1997) show that this emission is more 
widely spread than Murdin reported, suggesting that the material is not
associated with the Crab. We discuss here whether this material is 
associated with the wind from an O~star lying in the direction of the Crab.

We discuss these observations and present the data
in Section~2. In Section~3 we interpret the data with respect to the Crab
and a nearby O~star, and summarize the paper in Section~4.
                                                                              
\section{Observations and Results}

Figure~\ref{26mpics} reproduces images from WLT94, showing \ion{H}{1} emission around
the Crab Nebula with an angular resolution of $36'$. In that paper, three
approximately circular ``bubbles'' were identified in the \ion{H}{1} distribution.
The boundaries of these features are overlaid on the images in 
Figure~\ref{26mpics}: the larger bubble was denoted ``bubble~1'' and the 
middle one ``bubble~2.'' WLT94 associated the smallest bubble, which we will 
refer to here as ``bubble~3,''  with the O-star SAO77293; this association
is re-assessed below. We have also indicated the positions of the Crab
Nebula and SAO77293 with asterisks in Figure~1 (the lower asterisk denotes the
position of SAO77293).

\subsection{Effelsberg Observations}

The Effelsberg data image a $4.5^\circ$ square region centered on ``bubble~2'', 
indicated by the white box in Figure~\ref{26mpics}.
Details of the observations can be found in Table~\ref{Efftable}.
The data were taken during the winter of 1993/1994, during periods 
of weather too poor for observations
at higher frequencies; the resulting data are thus a patchwork of
several observations. Individual spectra were baseline
corrected using a third order polynomial and were adjusted for stray radiation
(Kalberla et al., 1982) before conversion to channel maps. Baselines were
determined using data between $-170$ and $-70$~km/s and $50$ to 130~km/s.

Due to the strong absorption of the Crab Nebula, the baseline fitting 
algorithm failed for spectra in a $\sim 25'$ region
on and immediately around the
Crab Nebula, rendering the data in that area inadequate for
addition to the DRAO data (see below). The data for
this region were excised and replaced with
values estimated from the surrounding emission. The estimated values were
derived by fitting a ``twisted-plane'' (a plane which varies linearly in
intensity along any cut in ``x'' or ``y'') 
to the emission immediately around the excised region. 
The Effelsberg data are intended to image the large-scale structure around
the Crab, and this process does not affect our interpretation of these
structures.

The final channel maps include a small amount of striping, a
result of calibration differences in data collected
at different epochs. Some channels also have pixels with unusually high
values compared to adjacent pixels; these defects pose no problems 
to the present work and have been left in the data. The r.m.s. 
noise in the maps is 0.25~K.

Plots of the Effelsberg data are presented in Fig.~\ref{corrected}. There is
a significant gradient in the level of emission across the field of view, so a
twisted-plane background has been fit to edges of the field and
removed from each channel for presentation purposes.
The lack of velocity dispersion towards the Galactic anti-center implies that
\ion{H}{1} associated with the Crab Nebula may be found in almost any velocity
channel, and we present a wide range of velocity channels for the sake of
completeness. Nevertheless, we discuss only those features which we consider 
to be in the vicinity of the Crab Nebula and the O star which lies near it in
the sky.

\subsection{DRAO Synthesis Telescope Observations}

Two overlapping fields were observed using the DRAO Synthesis telescope
(Roger et al., 1973; Veidt et al., 1985). 
One field is centered on  the position of the Crab,
while the second is centered approximately $1^\circ$ to the southeast near
the center of a \ion{H}{1} bubble which overlaps the position of the
Crab Nebula.
The details of these observations are presented in Table~\ref{DRAOobs}. 

The continuum contribution to the data was estimated by averaging data
between $22.8$ and $35.0$~km/s, and between $-32.6$ and $-57.4$~km/s to
create two end-channel continuum maps.
The strong emission from the Crab Nebula, coupled with 
non-linearities in the observing system (which are not important when observing
weaker sources), result in artefacts which vary with velocity from map to map.
Individual continuum maps were created for each channel by
linearly interpolating between the two extreme continuum maps
in order to subtract the most
reasonable estimate of the continuum emission (complete with artefacts) from
each channel map.  The interpolated channel maps were 
scaled by the Crab Nebula's integrated absorption in each channel before subtraction to
avoid over-subtraction of artefacts related to the SNR in channels
where the Crab Nebula is strongly absorbed.

The resulting maps reveal a large amount of \ion{H}{1} structure, but imaging
artefacts remain. In particular, the field centered on the
Crab Nebula has noticeable residual rings centered on the
SNR. Greisen (1973) showed that \ion{H}{1} absorption 
is not constant across the face of the
Nebula, and that the absorption varies with velocity. The continuum subtraction
process works correctly only if the absorption
across the face of the nebula is constant and as a result this process
improperly
subtracts the continuum component (and, more importantly, associated artefacts)
from our data.
An attempt was made to remove the effects of this non-uniform absorption by
CLEANing both the original and continuum-subtracted maps, but
there was only limited improvement over the uncleaned  maps,
possibly because of non-linearities in the receiver system.
Nevertheless, the data used here for the field centered on the Crab
have been CLEANed.  No CLEANing was necessary for data from
the southern field; the residual grating rings in this field
are below the noise in the map, because primary beam
attenuation of the Crab's emission has 
reduced the effects of the non-uniform absorption across the Crab.

The DRAO Synthesis Telescope is not sensitive to 
structures on size scales greater than $\sim 1^\circ$ at 21cm. To portray
the full range of \ion{H}{1} structure it is necessary to add single-antenna
information, in this case the data from the Effelsberg radio telescope presented
above, to the DRAO data.
The short and long-spacing datasets were filtered in complementary 
fashion in the uv-plane and, 
after correction for the primary beams of the two instruments, added in the
image plane.  The results of this process are entirely trustworthy for
regions $> 25'$ from the Crab Nebula. For regions closer to the Crab
the absolute flux may be affected by the interpolation of the Effelsberg
data in this region, discussed above,
but the morphology of the small scale emission is trustworthy.
The DRAO data (without short-spacings added) were consulted when
interpreting emission in the region immediately around the Crab
Nebula.

After the short spacing data had been added to each of the DRAO fields, the
data were re-projected to a common grid and mosaiced to form the
final data cube. The noise level near the center of the
southern field is 0.9~K. The residual rings seen in the northern field
are up to 20~K in magnitude and decrease in strength with distance from the
Crab.

The final
DRAO image data are presented in Fig.~\ref{draodata}. A constant background, 
determined from the average level of emission near the center of the observed
region, has been removed from each channel for presentation purposes.

\section{Discussion}

\subsection{Does the Crab Nebula lie in a void?}

Based on their low resolution \ion{H}{1} maps, WLT94
conclude that the Crab Nebula
lies in a low-density void which they label ``bubble~2.''
The higher resolution images presented here allow us to re-examine
this conclusion and study in more detail the environment of the Crab Nebula.

The lack of velocity dispersion in the Galactic
anti-center direction makes assigning a systemic velocity to the Crab
Nebula a difficult task.
The velocity crowding has the effect that even small non-circular motions
in the gas in front of the Crab could give rise to absorption at velocities
unrelated to the SNRs kinematic distance. 

In Figure~\ref{abs_spec} we present an absorption spectrum for the Crab
Nebula derived from the DRAO field centered on the Crab.
For comparison we include an emission spectrum obtained from the combined
Effelsberg and DRAO data for this same field. The emission spectrum was 
derived for an elliptical annulus around the Crab, the inner edge
having major and minor axes of $32'$ and $16'$ respectively, and the
outer edge have major and minor axes of $42'$ and $26'$; the major axis
of the ellipse is oriented along the declination axis. The inner edge of
the ellipse was chosen to avoid contamination from the Crab Nebula and
artefacts associated with it, while the choice of outer edge was arbitrary. 
The large size of the annulus will undoubtedly allow contamination of the
emission spectrum from gas some distance away from the Crab, and will thus
potentially over-estimate the emission at the position of the Crab itself.

The emission spectrum shows a number of features. There are two, strong,
emission peaks centered near $4$ and $-11$~km/s, a weak plateau near
$-22$~km/s, and a faint tail which reaches zero near $-52$~km/s (the
velocity at which the emission reaches zero may be affected by our choice of
channels used to create the end-channel continuum map at the
negative end of the spectrum).
The absorption spectrum reveals two major features, one centered near
$11$~km/s and one centered near $3$~km/s. Both absorption features
lie within the velocity range covered by emission associated with the feature
at $4$~km/s. There is no strong absorption associated with the
emission feature centered near $-11$~km/s. There is some indication of
a slight slope to the baseline of our absorption spectrum, and we are unable to
comment on the weak absorption measured by Hughes et al. (1971).

Both
Greisen (1973) and Radhakrishnan et al. (1972) fit the absorption to
the Crab with multiple gaussian components but neither study attempted
to fit the weakest absorption. Radhakrishnan et al. (1972) fit the
last component at $-1.5$~km/s, while Greisen fits one at $\sim -8$~km/s, but
weak \ion{H}{1} absorption of the Crab's continuum
emission persists to very high negative velocities (e.g. Hughes et al., 1971).
The component at $-8$~km/s fit by Greisen appears to be an attempt to 
account for this weak absorption, and for the non-gaussian shape of the deep
absorption feature centered at $\sim 3$~km/s. An absorption spectrum derived
from our DRAO data shows that, at $-8$~km/s, the absorption has 
declined smoothly to less than 1\% of the continuum level; 
we do not feel that this absorption is significant. 

Inspection of Figure~\ref{draodata} reveals that the gas giving rise to the 
absorption peak centered at $\sim 3$~km/s is part of the bright western rim of 
bubble~2, which extends in velocity to $\sim 1$~km/s. The Crab Nebula
must lie behind this gas. At $-7$~km/s a bar of emission
crosses the position of the Crab; there is no indication of absorption of
the Crab's emission at this velocity, and this gas must lie behind the
Crab. These arguments allow us to place the Crab between these two features
in velocity, and the systemic velocity of the Crab must therefore
lie in the range $1$ to $-7$~km/s.

In the following discussion, velocities quoted for various features are those 
derived from the data at full resolution. Figures \ref{corrected} and 
\ref{draodata} present images at a velocity separation of
2~km/s and maps at the exact velocities quoted in the text may not appear. 
All features can, however, be seen in the data presented.

In Figure~\ref{26mpics}, bubble~2 is seen
from $v_{\rm LSR}=5.8$~km/s to $v_{\rm LSR}=-17.2$~km/s, is roughly
$3.8^\circ$ in diameter, and is centered at $\alpha = 5^h 37^m 36^s$,
$\delta = 21^\circ 47' 44"$. In the Effelsberg data bubble~2 is seen
very clearly at $-3.3$~km/s and $-4.6$~km/s, 
centered near $\alpha \sim 5^h 38^m 20^s$, $\delta \sim
21^\circ 50'$, and has a diameter of $\sim 3^\circ$. Although the structure is
somewhat obscured by confusion, the diameter clearly decreases towards
more positive velocities.
It is first seen as a very faint shell at $7.8$~km/s, and grows
in prominence as velocity decreases. At 5.3~km/s a prominent arc of emission
defines the western edge of the bubble and at $1.2$~km/s there is a prominent
arc to the east. The western arc
disappears by $-0.5$~km/s at which point
the bubble is defined largely by the
semi-circular bay to the west. The
eastern edge merges with the bright emission to the NE at $-2.1$~km/s and a
faint arc to the south of the field bounds the structure in that
direction. At $-3.8$~km/s the void is at its
maximum size and by $-6.2$~km/s is no longer recognizable in the
Effelsberg data. At more negative velocities, the \ion{H}{1} peak located near
the geometrical center of bubble~2 at
$-6.2$~km/s eventually grows into a ridge which
defines one edge of the \ion{H}{1} bubble centered at $v_{LSR} = -12$~km/s.

Note that the uneven \ion{H}{1} shell that bounds bubble~2 at $-7.3$~km/s 
in WLT94 is
not seen in the Effelsberg data. If the shell exists at and beyond this 
velocity, it 
lies just at the edge of the field of view. For the purposes of this
paper we use the velocity width derived from the Effelsberg data, 7.0 to 
$-6.0$~km/s.

We can identify bubble~2 over a velocity range of 13~km/s, roughly twice the
width of the average ISM feature. Furthermore, bubble~2 changes its diameter 
systematically with velocity. We conclude that the bubble is expanding (or
contracting) and that its width in velocity is not simply a product of
turbulent motion. We take the expansion velocity to be $\sim 6.5$~km/s.
The average brightness temperature of the shell (above the background)
is $3.8 \pm 0.6$~K, and the mass of the shell is
$4 \times 10^4$~M$_\odot$ (assuming helium is 1/10 as abundant as
hydrogen by number and that the
\ion{H}{1} shell is at the same distance as the Crab Nebula (2~kpc)). 
The energy in the shell is $\sim 2 \times 10^{49}$~ergs and
the kinematic age is $8 \times 10^6$~yrs.

The Effelsberg data show that there is \ion{H}{1} structure
projected inside bubble~2 in the range of systemic velocities for
the Crab.  The resolution of the Effelsberg data is insufficient
to investigate the immediate surroundings of the Crab, however. For this
reason the DRAO \ion{H}{1} data in this velocity range were examined for
evidence of either a void or an interaction; these data are discussed below.

At $v=3.7$~km/s in Figure~\ref{draodata} the Crab Nebula
is seen near the inner edge of the bright western rim of bubble~2. As
velocity decreases the rim becomes less prominent, but \ion{H}{1} emission
still remains around the position of the Crab. From $v \sim 3$ to $v \sim
-5.5$~km/s there appears to be very little \ion{H}{1} immediately
around the position of the Crab, but by $v = -7.0$~km/s
the bar of emission which defines the SW edge of the bubble centered at
$v_{LSR} = -12$~km/s
has crossed the position of the Crab.

The absorption in this range is quite weak 
so it is difficult to decide whether the emitting
material from $v \sim -3$ to $v \sim -5.5$~km/s is in front
of, behind, or surrounding the Crab Nebula. If this (or other)
material were actually surrounding the Crab one might expect to
see evidence of an interaction (e.g. a swept-up shell).
The next question is whether there is any evidence of
an interaction between the Crab and the \ion{H}{1}. There are in fact two
possibilities, based largely upon positional coincidences between
the \ion{H}{1} structures and the Crab.

The first appears at $v= -7.0$~km/s.  At this velocity
there is some \ion{H}{1} emission, in a rough semi-circular
arc, to the NW of the SNR. By $v=-9.5$~km/s
this arc has encircled the entire top of the SNR, with a small
intensity depression just above the position of the Crab
mirroring an intensity peak just below the Crab.
Errors in the phase calibration of radio interferometer data often
take the form of asymmetric artefacts in the resulting images,
and the peak and depression are, in our opinion, residual mapping
and calibration artefacts.  The emission to
the NW is the brightest, and by $v=-12.0$~km/s the
emission to the NE has disappeared. The emission immediately
around the Crab has disappeared totally by $v=-16.1$~km/s.

The other indication of a possible interaction is first seen immediately 
to the east of the Crab at $v=-10.4$~km/s.
It is manifested as a thin arc of emission which appears to
encompass the entire eastern edge of the SNR out to $v=-15.3$~km/s.
The structure lies within and along the contours of the
Crab Nebula, and may be an artefact caused
by imperfect removal of the Crab's continuum emission, possibly due to the
changing absorption across the face of the Crab.

We conclude that there is no strong evidence of an interaction between the Crab
Nebula and any \ion{H}{1} in its vicinity, and confirm the conclusions
of WLT94 -- that the Crab Nebula lies in the low-density region identified as
``bubble~2.''
The features which may be taken as evidence for an interaction
between the Crab Nebula and its surroundings
are explained as simple map artefacts resulting from the
extreme brightness of the Crab Nebula.  This does not mean that there
is no interaction, only that the effects of any such interaction are below
the level of the artefacts in the present data.
Further observations, capable
of dynamic range greater than 400:1 which we have achieved, 
are required to settle this question.

\subsection{A remnant stellar wind?}

Murdin (1994) reports H$\beta$ and optical continuum observations which
he interprets as revealing an extended halo of stellar wind material around the
Crab Nebula. He attributes this wind material to the Crab Nebula's progenitor.
Fesen et al.~(1997) shows that the H$\beta$ emitting material is more widely
distributed than Murdin (1994) reports, suggesting that this material is
unrelated to the Crab. In this section we show that a stellar wind bubble
exists along the line of sight towards the Crab, and hypothesize that the
emitting material found by Murdin~(1994) and Fesen et al.~(1997) is associated
with this bubble and not the Crab Nebula.

The new observations confirm the interpretation of WLT94 that there is an 
expanding \ion{H}{1} shell, which bounds the feature we will call
``bubble~3,'' centered to the south of the Crab Nebula,
but show that this bubble is {\it not} powered by the O7.5 star
SAO77293 as suggested in that paper. The feature which we interpret 
as this shell is clearly seen in both the Effelsberg and 
DRAO \ion{H}{1} data (Figures~\ref{corrected} and \ref{draodata}). 
There are two parts to the shell. From
$-6$ to $-12$~km/s the apparent radius increases from almost zero to
$50'$. Between $-12$ and $-14$~km/s the radius
suddenly increases to $60'$ and continues to increase to a velocity of
$-25$~km/s, where the shell fades from view having a radius of $75'$. 
This suggests an expanding shell
in two parts with a systemic velocity near $-12$~km/s, with the
red-shifted part expanding into a more dense medium while the blue-shifted
part expands into a less dense region. 

The brightest portion of the \ion{H}{1} shell is its NW sector, where
it joins a complex of bright emission running to the NW between $-15$ and 
$-25$~km/s (it should be noted that
this area includes the position of the Crab Nebula, and is the area where
the Effelsberg data were interpolated). It is
unclear whether the northern bright emission
is physically associated with the shell or whether it is an unrelated
result of velocity crowding. We therefore make two calculations
of the \ion{H}{1} mass, one including the
bright emission and one avoiding it. The calculations
are made by integrating the flux within a polygon around the shell
and subtracting a baseline determined by fitting a twisted plane to the
vertices of the polygon. Errors are estimated by making several measurements
using different polygons and taking the standard deviation of the resulting
values. Including the bright emission to the north, the
average brightness temperature over the velocity range $-6$ to
$-26$~km/s is ${\rm T_B}=4.9 \pm 0.4$~K;
excluding it yields ${\rm T_B}=3.5 \pm 0.4$~K. Using these
estimates the mass of
the \ion{H}{1} shell is $(8.7 \pm 0.6) \times 10^3 {\rm M}_\odot$ and
$(4.1 \pm 0.6) \times 10^3 {\rm M}_\odot$, respectively.

A stellar wind bubble in a homogeneous medium with number density
$n_\circ$ (Weaver et al., 1977) expands such that its radius,
$r_b$, at a given time is
\begin{equation}
r_b=0.763\bigl({\dot{E}\over{1.4 m_h n_\circ}}\bigr)^{1/5} t^{3/5}
\label{first-eqn}
\end{equation}
where $\dot{E}$ is the mechanical luminosity of the wind 
($= 1/2 \dot{M} v^2$), $m_h$ is the mass of a hydrogen atom,
and $t$ is the age of the bubble. The bubble 
expansion velocity, $v_b$, is found by taking the time 
derivative of its radius, giving
\begin{equation}
v_b = 0.6 r_b /t.
\label{second-equation}
\end{equation}

We can estimate the mechanical luminosity of the stellar wind 
by first estimating the size, age
and pre-swept density of the bubble.
As noted earlier, bubble~3 has two distinct size scales; 
we attribute this 
to a ``blow-out'' of the bubble into a lower density region.
Using the size scale of the smaller region ($50' = 29$ pc at 2~kpc),
and taking the expansion velocity of the bubble to be 11~km/s (half of its
velocity width), we use Equation~\ref{second-equation} to estimate the
age of the bubble as $1.5 \times 10^6$~years.
Finally, if we assume that the entire mass of \ion{H}{1} was
evenly distributed in a volume corresponding to the volume of the smaller
shell, the undisturbed ISM density was between $\sim 1.6$ and $3.5$~cm$^{-3}$,
for the lower and higher mass respectively. 
Substituting these values into Equation~\ref{first-eqn} we estimate
the mechanical luminosity of the powering wind to be $\sim 10^{36}$~erg/s,
suggesting that an O6 ($L_w \sim 1.8 \times 10^{36}$ erg/s) or O7 
($L_w \sim 6.8 \times 10^{35}$ erg/s) type star is at the 
heart of the bubble (assuming
the empirical relationships for mass loss rate and terminal velocity derived
by Howarth and Prinja, 1989).

The kinetic energy of the \ion{H}{1} shell
is $4.5  - 7.8 \times 10^{48}$~ergs, much less than $\sim 1.4 \times 
10^{50}$~ergs that an O6 star would inject into the ISM over its main-sequence
lifetime of $\sim 4.2 \times 10^6$~years.  The amount of energy lost through
recombination of the material in the swept-up shell can be estimated as 
the number of hydrogen atoms in the
shell times the ionization potential; this comes to $\sim 10^{50}$~ergs
given the mass estimates above.
For the parameters derived or estimated
here, we find that the ionization front associated with an O6 or O7 star
would become trapped after $\sim 2 - 4 \times 10^6$~years, depending on the
pre-swept density of the ISM; this is consistent with the estimated age
of the \ion{H}{1} bubble. These results
indicate that the \ion{H}{1} shell
is due to a trapped ionization front associated with the stellar wind bubble.

The only known O-star along the line of sight to the 
\ion{H}{1} bubble is SAO77293, which WLT94
associate with the \ion{H}{1} bubble. This star, at an estimated
distance of 1.8~kpc, is of type O7.5III. Despite the rough agreement in
stellar type with our estimates above, and distance to the bubble, we
suggest that SAO77293 is {\it not} associated with bubble~3. 
Christy~(1977) measured
an absorption feature due to the interstellar \ion{Ca}{2} K line in the spectrum
of SAO77293 at a velocity of $-11 \pm 3$~km/s, placing the star behind
the gas at this velocity. Our \ion{H}{1} maps show a region of bright emission, at the
position of SAO77293, from $1$ to $\sim -13$~km/s; we associate the gas giving rise to
the \ion{Ca}{2} K-line absorption with this complex of \ion{H}{1}.
The emission at these velocities is part of
the red-shifted portion of the \ion{H}{1} shell,
and if the \ion{H}{1} shell is expanding and not contracting, SAO77293 must be located on the
far-side of the \ion{H}{1} bubble in order for the \ion{Ca}{2} K line absorption
to exist. If SAO77293 were to lie within the gas producing the \ion{Ca}{2} K line absorption,
the radiation from the star would ionize the surrounding gas; since we see no evidence
for a hole in the \ion{H}{1} distribution at these velocities we rule out this
possibility. Finally, if the star were to lie at a systemic velocity at $-14$~km/s or
beyond, it would clearly lie within the confines of the blue-shifted, blow-out, region
of bubble~3, and would be in front of the gas giving rise to the \ion{Ca}{2} K line absorption.
This means that SAO77293 must be behind the \ion{H}{1} 
bubble and cannot be powering it, and we conclude that the star which
powers the \ion{H}{1} bubble has yet to be identified. 
We are then left with the unanswered question, why has that star not been
detected?

The Crab lies projected on the edge of the \ion{H}{1} shell defining the
red-shifted portion of bubble~3, and within the projected
confines of the blue-shifted portion. The stellar wind material within the
\ion{H}{1} shell is likely ionized and can give rise to 
H$\beta$ emission. In addition, if the gas in the \ion{H}{1} shell is
not fully recombined we can expect some additional H$\beta$ emission
from the shell itself. We thus suggest that the H$\beta$
emission detected by Murdin~(1994) and Fesen et al.~(1997) is unrelated to
the Crab Nebula and is, instead, due to material
associated with the (as yet unidentified)
star which is powering bubble~3.

\section{Summary}
In this paper we have presented maps of the \ion{H}{1} environment of the
Crab Nebula obtained using the DRAO ST and the Effelsberg 100~m Radio 
Telescopes. The Crab Nebula lies within the physical confines of the 
\ion{H}{1} structure which we call bubble~2. This association is based upon the
observation that the central velocity of the \ion{H}{1} bubble lies within
the range of possible systemic velocities of the Crab Nebula.
A few maps appear to show interaction, but these features are interpreted as
artefacts resulting from the extreme brightness of
the Crab Nebula. 

We also confirm the presence of an expanding \ion{H}{1} bubble (labelled
``bubble~3'')  along the line of sight to the Crab Nebula. The emission from
the Crab is not absorbed by the \ion{H}{1} shell associated with the bubble;
bubble~3 must therefore lie behind the Crab. 
We conclude that an association between
bubble~3 and the nearby O-star SAO77293 is unlikely, and that bubble 3 lies
in front of this star. It is suggested 
that emission from material associated with bubble~3 has been detected by
Murdin~(1994) and Fesen et al.~(1997).

\section{Acknowledgements}

The Dominion Radio Astrophysical Observatory
is operated as a national facility by the National Research Council of Canada.
This research was supported in part by a grant from the Natural Sciences
and Engineering Council of Canada.

\label{sec-obs}
%\ion{H}{1}

%\clearpage

\vfill\eject

\begin{figure}
\caption[26mpic.ps]{Low-resolution, wide-field image of the \ion{H}{1}
distribution around the Crab Nebula from WLT94. A twisted-plane background
has been removed from each frame for presentation purposes. The large, medium,
and small circles denote the boundaries of the features ``bubble~1,'' 
``bubble~2'' and ``bubble~3.''. The upper asterisk marks the position of the 
Crab Nebula, while the lower marks 
the position of SAO77293. The white box outlines the approximate area covered
by the Effelsberg 100-m radio telescope presented in this paper. The greyscale
runs from $-10$ to 10~K, with contours every 3~K. More
details on these data can be found in WLT94.}
\label{26mpics}
\end{figure}

\begin{figure}
\caption[test1.ps]{The Effelsberg \ion{H}{1} data for the region around the
Crab Nebula are shown in greyscale with overlayed contours; the contours and
greyscale limits are different for each page and are given below. The
position of the Crab Nebula (upper) and SAO77293 (lower) 
are given by the small white
circles in the channel maps. A twisted-plane background 
has been subtracted from
each channel for presentation purposes. The greyscale runs 
from $-2.3$ (white) to 9.3~K
(black) in (a), $-14.8$ to 21.2~K in (b), $-23.0$ to 32.3 in (c), $-28.2$
to 34.3 in (d), $-13.4$ to 21.8 in (e), $-4.5$ to 9.5 in (f), $-1.8$ to
5.4 in (g), and $-1.4$ to 2.2 in (h).
Contours appear approximately
every 0.7~K in (a), 2.0~K in (b), 3.1~K in (c), 3.5~K in (d), 2.0~K in (e),
0.8~K in (f), 0.4~K in (g), and 0.2~K in (h).}
\label{corrected}
\end{figure}

\begin{figure}
\caption[test2.ps]{The DRAO \ion{H}{1} data for region around the Crab Nebula are
shown in greyscale; the limits of the greyscales are different for each page
and are given below. The Crab Nebula is
located in the center of the upper field, at the center of the artefacts.
The greyscales run from $-27.5$ (white) to 24.5~K (black) in figure (a),
and $-26.2$ to 25.6~K in figure~(b).
}
\label{draodata}
\end{figure}

\begin{figure}
\caption[ems+optd.ps]{The absorption spectrum (dashed line) of, and an 
emission spectrum (solid line) towards, the Crab nebula. The scale for the
emission spectrum is on the left-hand axis, while the scale for the
absorption spectrum is found on the right. Details of the 
spectra can be found in the text.}
\label{abs_spec}
\end{figure}

\tablenum{1}
\tablewidth{0pt}
\begin{deluxetable}{l c c}
\tablecaption{Effelsberg Observation Parameters}
\startdata
Field Center & RA(J2000) & 05$^h$ 37$^m$ 35.55$^s$ \\
& Dec(J2000) & 21$^\circ$ $47'$ $30"$\\
Field Size & & $4.5^\circ \times 4.5^\circ$ \\
Scan Length & & $4.5^\circ$ \\
Scan Separation & & $4.5'$ \\
Spatial Resolution  & & $9.4' \times 9.4'$\\
Central Velocity & & $-10$~km/s\\
Total Bandwidth & &1.5625 MHz\\
Velocity Resolution & & 0.644~km/s \\
Integration Time & & 30s/30s (on/off frequency)\\
\enddata
\label{Efftable}
\end{deluxetable}

\tablenum{2}
\tablewidth{0pt}
\begin{deluxetable}{l c c}
\tablecaption{DRAO Observation Parameters}
\startdata
Field Center & RA(J2000) & 05$^h$ 37$^m$ 19$^s$ \\
& Dec(J2000) & 21$^\circ$ $00'$ $00"$\\
Field Center & RA(J2000) & 05$^h$ 34$^m$ 30$^s$ \\
& Dec(J2000) & 22$^\circ$ $01'$ $00"$\\
Field Size (to 10\%) & & 3.1$^\circ$\\
Spatial Resolution  & & $1.0' \times 2.75'$\\
Central Velocity & & $-12$~km/s\\
Total Bandwidth & &0.5 MHz\\
Velocity Resolution & & 1.3~km/s\\
Channel Separation & & 0.824~km/s\\
\enddata
\label{DRAOobs}
\end{deluxetable}

\tablenum{3}
\tablewidth{0pt}
\begin{deluxetable}{l c c c c c}
\tablecaption{Observed characteristics of features and objects}
\startdata
%& Bubble 1 &Bubble 2 &Bubble 3 &Crab Nebula &SAO77293\\
%Center RA & $5^h\ 36^m$& $5^h\ 37^m\ 36^s$ &$5^h\ 35^m\ 48^s$&$5^h\ 34^m\ 29.8^s$&  $5^h\ 35^m\ 39.8^s$ \\
%Center Dec & $20^\circ\ 53'$&$21^\circ\ 47'\ 44"$ &$21^\circ\ 52'\ 50"$&$22^\circ 1'$& $21^\circ\ 23'\ 58.0"$\\
%Maximum size & $6.4^\circ$& $3.8^\circ$ & $50' - 75'$ &$7' \times 5'$& -- \\
%Velocity range (km/s) & $(9.2,-16)$ & $(5.8,-17.2)$& $( -6,-25)$& $(1,-7)$ & $< -11 \pm 3$\\
%Distance (kpc) & & & & 1.5-2.3 & $\sim 1.8$\\
Name & Center RA & Center Dec & Maximum size & Velocity range& Distance \\
& (J2000)& (J2000)& & (km/s) & (kpc)\\
Bubble 1 & $5^h\ 36^m$& $20^\circ\ 53'$& $6.4^\circ$& $(9.2,-16)$&\\
Bubble 2 & $5^h\ 37^m\ 36^s$ & $21^\circ\ 47'\ 44"$ & $3.8^\circ$ & $(5.8,-17.2)$&\\
Bubble 3 & $5^h\ 35^m\ 48^s$& $21^\circ\ 52'\ 50"$& $50'-75'$ & $(-6,-25)$&\\
Crab Nebula & $5^h\ 34^m\ 29.8^s$& $22^\circ 1'$& $7' \times 5'$& $(1,-7)$ & 1.5-2.3\\
SAO77293 & $5^h\ 35^m\ 39.8^s$& $21^\circ\ 23'\ 58.0"$& -- & $< -11 \pm 3$& $\sim 1.8$\\
\enddata
\label{feature_tab}
\end{deluxetable}

\vfill\eject

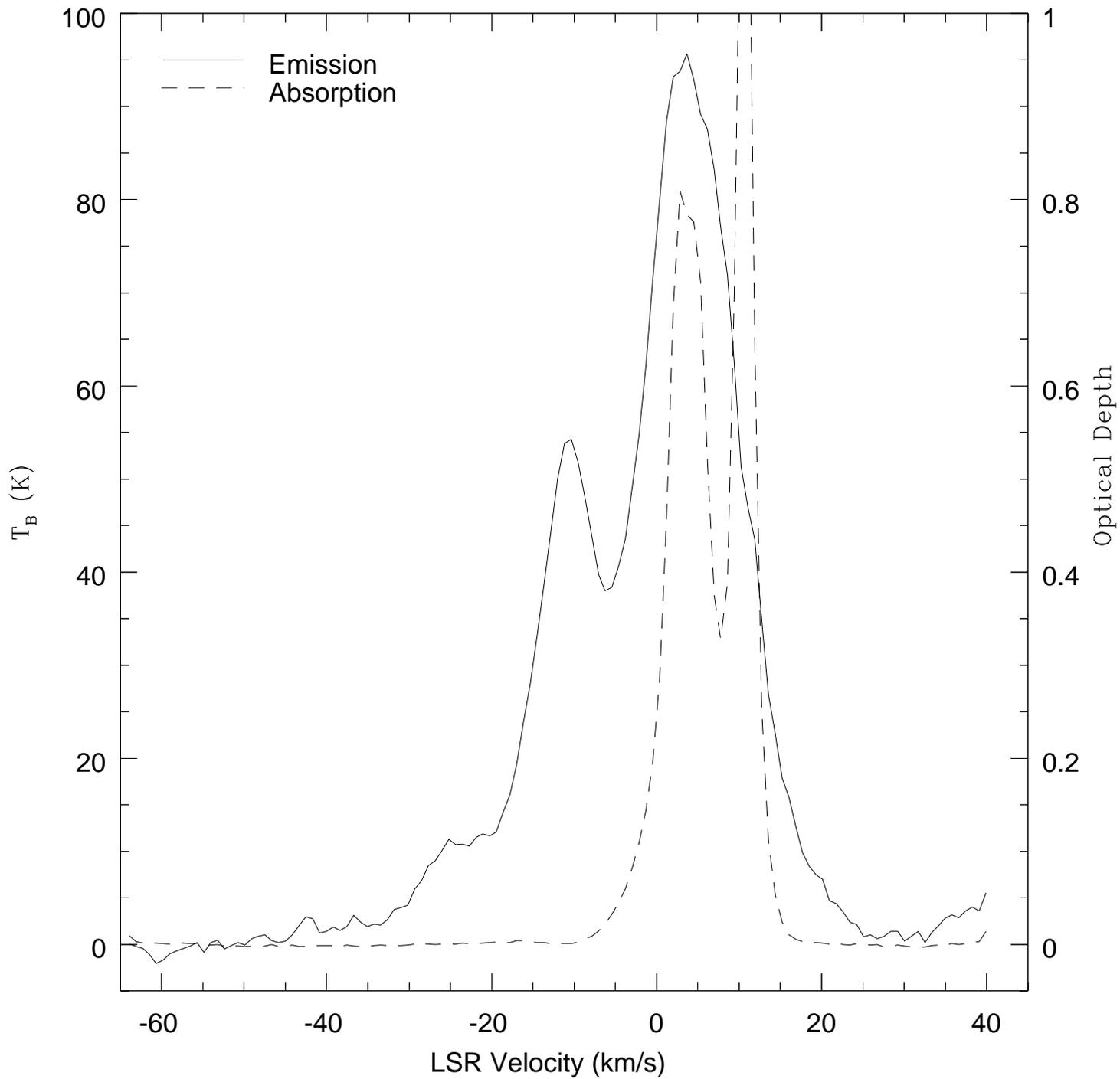
\end{document}